\documentstyle[11pt,aaspp]{article}
\input epsf
\def\l*{$L_*$\/}
\def\etal{{\it et al. }}

\def\kms{\rm ~km~s^{-1}}

\def\kmsmpc{km s$^{-1}$ Mpc$^{-1}$\ }

\def\l*{$L_{*}$}

\def\gsim{ \lower .75ex \hbox{$\sim$} \llap{\raise .27ex \hbox{$>$}} }
\def\lsim{ \lower .75ex \hbox{$\sim$} \llap{\raise .27ex \hbox{$<$}} }
\def\pp{\noindent\parshape 2 0truecm 16.0truecm 2.0truecm 15truecm}

\def\spose#1{\hbox to 0pt{#1\hss}}
\def\simlt{\mathrel{\spose{\lower 3pt\hbox{$\mathchar"218$}}
     \raise 2.0pt\hbox{$\mathchar"13C$}}}
\def\simgt{\mathrel{\spose{\lower 3pt\hbox{$\mathchar"218$}}
'     \raise 2.0pt\hbox{$\mathchar"13E$}}}

\font\titlefont=cmss17

\input{epsf}

\begin{document}

\title{\titlefont Cold collapse and the core catastrophe}


\vskip 0.5truecm


\centerline{\bf B. Moore$^{\bf 1}$, 
T. Quinn$^{\bf 2}$, 
F. Governato$^{\bf 3}$,
J. Stadel$^{\bf 2}$, 
G. Lake$^{\bf 2}$
}

\

\parskip=0pt

$^{\bf 1}$ {Department of Physics, University of Durham, Durham City, DH1 3LE, UK}

$^{\bf 2}$ {Department of Astronomy, University of Washington, Seattle, WA 98195, USA}

$^{\bf 3}$ {Osservatorio Astronomico di Brera - Merate, Italy}

\parskip=7pt

\

\begin{abstract}

We show that a universe dominated by cold dark matter fails to reproduce the rotation
curves of dark matter dominated galaxies, one of the key problems that it was designed to
resolve.  We perform numerical simulations of the formation of dark matter halos, each
containing $\gsim 10^6$ particles and resolved to 0.003 times the virial radius, allowing
an accurate comparison with rotation curve data.  A good fit to both galactic and cluster
sized halos can be achieved using the density profile $\rho(r) \propto
[(r/r_s)^{1.5}(1+(r/r_s)^{1.5})]^{-1}$, where $r_s$ is a scale radius.  This profile has a
steeper asymptotic slope, $\rho(r)\propto r^{-1.5}$, and a sharper turnover than found by
lower resolution studies.  The central structure of relaxed halos that form within a
hierarchical universe has a remarkably small scatter (unrelaxed halos would not host
disks).  We compare the results with a sample of dark matter dominated, low surface
brightness (LSB) galaxies with circular velocities in the range $100\--300\kms$.  The
rotation curves of disks within cold dark matter halos rise too steeply to match these
data which require a constant mass density in the central regions.  The same conclusion is
reached if we compare the scale free shape of observed rotation curves with the simulation
data.  It is important to confirm these results using stellar rather than HI rotation
curves for LSB galaxies.  We test the effects of introducing a cut-off in the power
spectrum that may occur in a universe dominated by warm dark matter. In this case halos
form by a monolithic collapse but the final density profile hardly changes, demonstrating
that the merger history does not play a role in determining the halo structure.

\end{abstract}

\vfil\eject

\section{Introduction}

Determining the nature of dark matter remains the most important unsolved problem in
modern cosmology. Most baryonic candidates have been ruled out by a variety of dynamical
and observational constraints, leaving a host of hypothetical particles as the most
popular contenders.  Of these, cold dark matter (CDM) remains the primary candidate and
provides a highly successful cosmogonic scenario (Davis \etal 1985).  
The basic premise is an
inflationary universe dominated by a dark matter particle, such as the neutralino or
axion, that leads to ``bottom up'' hierarchical structure formation. Small dense halos
collapse at high redshifts and merge successively into the large virialised systems that
can allow gas to cool and form stars and galaxies. The cold dark matter model was originally
motivated as a means for providing the mass fluctuation spectrum necessary to form
galactic halos and to explain the large scale clustering properties of galaxies 
(Peebles 1984).

A great deal of computational work has been devoted to resolving the structure of halos
that form within the cold dark matter model in order to facilitate a comparison with
observational data.  The first simulations of Quinn \etal (1986) and Frenk \etal (1988)
did not have the resolution necessary to probe the inner regions of dark halos that could
be compared with galactic rotation curves. However, it was encouraging for the CDM model
that the halos gave flat rotation curves beyond the resolution length of $\sim
50$ kpc.  Higher resolution simulations by Dubinski \& Carlberg (1991), Warren \etal
(1992) and Crone \etal (1994) showed evidence for cuspy
density profiles that varied as $\rho(r) \propto r^{-1}$ in the central regions.

Navarro, Frenk \& White (1996, hereafter NFW) made a systematic study of CDM halo
structure over a range of mass scales. They found, albeit with a large scatter, that the
density profiles of halos has a universal form, uniquely determined by their mass; 
varying from $r^{-1}$ in the central regions, smoothly rolling over to
$r^{-3}$ at the virial radii. (The virial radius, $r_{vir}$, is defined as the radius of a
sphere that has a mean enclosed density of 200 with respect to the critical value).
Many authors have verified the results of NFW using simulations of a similar resolution
{\it i.e.} 5,000 - 20,000 particles per halo and force softening $\sim 0.01r_{vir}$.  
(e.g. Cole \& Lacey 1996, Tormen \etal 1996, Brainerd \etal 1998, Thomas \etal 1998, 
Jing 1999; etc).  

This limited resolution has lead to over-interpretation of the halo profiles on scales
probed by galactic rotation curves. As the resolution is increased, not only are higher
central densities achieved, but the slope of the central density profile increases
(Moore \etal 1998). This result was also indicated by high resolution simulations
of isolated halo collapses by Carlberg (1994) and Fukushige \& Makino (1997).
A large sample of well resolved, galactic mass halos analysed by Kravtsov \etal (1998), 
gave central profiles with slopes shallower than found by NFW, a result that is at 
odds with the conclusions of this paper. 

Most of these studies have found a large variance in the halo structural
parameters, depending upon the degree of virialisation.  The current situation is
therefore somewhat confusing. If the cosmological scatter between halos of the same mass
is large, then the Tully-Fisher relation must be somewhat fortuitous \--- especially in
dark matter dominated galaxies such as low surface brightness (LSB) galaxies (e.g.  Zwaan
\etal 1995, Eisenstein \& Loeb 1997, Avila-Reese \etal 1998).  
If the central profiles are steeper than the NFW
profile, as the highest resolution studies find, then the hierarchical models will have
problems reproducing the observed rotation curves of galaxies.

In this paper we carry out a series of simulations of individual CDM halos extracted from
a large cosmological volume and re-simulated at higher resolution.  These are the first
results for a reasonably large sample of galactic mass halos, each resolved 
with $\gsim 10^6$
particles and integrated to a redshift z=0.  Simulation techniques and parameters are
summarised in Section 2. We examine the structure of these halos and make a comparison
with the rotation curves of LSB and dwarf spiral galaxies in Section 3, summarising our 
results in Section 4.

\section{Numerical techniques and simulation parameters}
 
Initially we perform a simulation of a 100 Mpc cube of a standard CDM universe with
$\Omega=1$, normalised such that $\sigma_8=0.7$ and the shape parameter $\Gamma=0.5$ (a
Hubble constant of 50 \kmsmpc\ is adopted throughout). The initial simulation contains $144^3$
particles with a mass of $2\times10^{10}M_\odot$ and we use a force softening of 60 kpc.
The particle distribution is evolved using the parallel treecode, ``PKDGRAV'', that has
accurate periodic boundaries and a variable timestep criteria based upon the local
acceleration.  The code uses a co-moving spline softening length such that the force is
completely Newtonian at twice our quoted softening lengths. 

At a redshift z=0 we select six halos with circular velocities in the range $130\--230
\kms$ (total mass $\sim 10^{12}M_\odot$) that are to be re-simulated at higher resolution.
Over this range of circular velocities, the NFW simulations predict that the halo
concentration should vary by $\lsim 15 \%$.  The halos we chose lie outside of clusters
and voids and are typically within filamentary structures;  no pre-selection was made on
their merger histories. Two of these halos are in binary system, with kinematics and
environment that resemble the Local Group. Since the initial halos contain just a few
hundred particles, we do not know if they will be completely virialised by the final
epoch.

The particles that lie within a sphere of radius twice their final virial radii are traced
back to the initial conditions to identify the regions that are to be simulated at higher
resolution.  The power spectrum is extrapolated down to smaller scales, matched at the
boundaries such that both the power and waves of the new density field are identical in
the region of overlap, then this region is populated with a new subset of less massive
particles.  Zones of heavier particles are placed outside this region, which allows the
correct tidal field to be modeled from the entire cosmological volume of the initial box.

The particle mass in the high resolution regions is $2 \times 10^6M_\odot$ and the spline
force softening is set to 1 kpc.  The starting redshift is increased such that the initial
fluctuations are less than one percent of the mean density, typically $z = 100$, and we
then re-run the simulation to the present epoch.  Each halo requires $\sim 50,000$ T3E cpu
hours to evolve to the present day. The particles on the shortest timestep require of
order 50,000 individual steps (see Quinn \etal 1999, for further details of the
time-stepping criteria), each calculation requires about $10^{15}$ floating point 
operations. 
The final virial radii of the halos agree with the low-resolution
runs to within a few percent and even the positions of the largest substructure clumps is
accurately reproduced.

\section{Results}

We identify the halo centers using the most bound particle which agrees extremely well
with the center of mass, recursively calculated using smaller spherical regions.  Once the
halos have collapsed, {\it i.e.}  90\% of their mass is in place, the density profiles do
not evolve with time, demonstrating that relaxation does not effect our results even in
the central regions.  Although a great deal of substructure is evident in these
simulations, approximately $85\%$ of the mass within the virial radius lies
within a smooth background of particles.

All of the systems we study are completely virialised by a redshift
z=0.  The largest major accretion of a halo with mass $\gsim 25\%$ of the final halo's,
occurred over a crossing time before the final time.  The density 
profiles immediately after such an accretion
do show strong variations and departures from the final state.  However, the disk
of a spiral galaxy suffering such a large merger would be destroyed and by the time a new
disk could have reformed, the halo would have achieved a new equilibrium.  For this reason
we believe that it is important to compare reasonably 
well relaxed halos with rotation curve data.

Figure 1 shows the rotation curves of our six CDM halos calculated directly from the
particle data, $v_c(r)=\sqrt{(GM/r)}$.  For each halo we rescale the circular velocities
and length scales by the same factor, such that the peak value is a fiducial $200 \kms$.
Our ``average'' fiducial 
galaxy has a virial mass of $1.4\times 10^{12}M_\odot$, a virial radius
of 288 kpc at which point the circular velocity has fallen to $144 \kms$. 
(This scaling is not unphysical since the virial radius scales as the circular
velocity.)
Note that the central rotation curves of the simulated
halos are all very similar, however, 
the scatter in the total mass is larger. Once scaled to the
same peak rotational velocity, the values of the circular velocity at
the virial radius vary by 15\%, {\it i.e.} the total masses vary by $\sim 30$\%.

In order to compare with observational data we compile a sample of rotation curves of a
dozen LSB galaxies with peak velocities in the range $100\--300\kms$. These are taken from
de Blok \& McGaugh (1997) and Pickering \etal (1997). 
We treat these rotation curves in the same way as
our CDM data, {\it i.e.} we rescale lengths and velocities by the same factor to give a peak
rotation curve of $200 \kms$. In the cases where the peak rotational velocity has not been
reached, we adopt a truncated isothermal plus core density profile, 
and fit the resulting rotation curve to the data
in order to estimate the scaling factor.  
(This would correspond to $\alpha=2$, $\beta=3$ and $\gamma=0$ in equation 1.)
We ignore the contribution from the HI
gas and stars such that the halo contribution is maximised - including the effects of
dissipating baryons only makes the discrepancy between the data and the model worse.

\centerline{\epsfysize=4.0truein \epsfbox{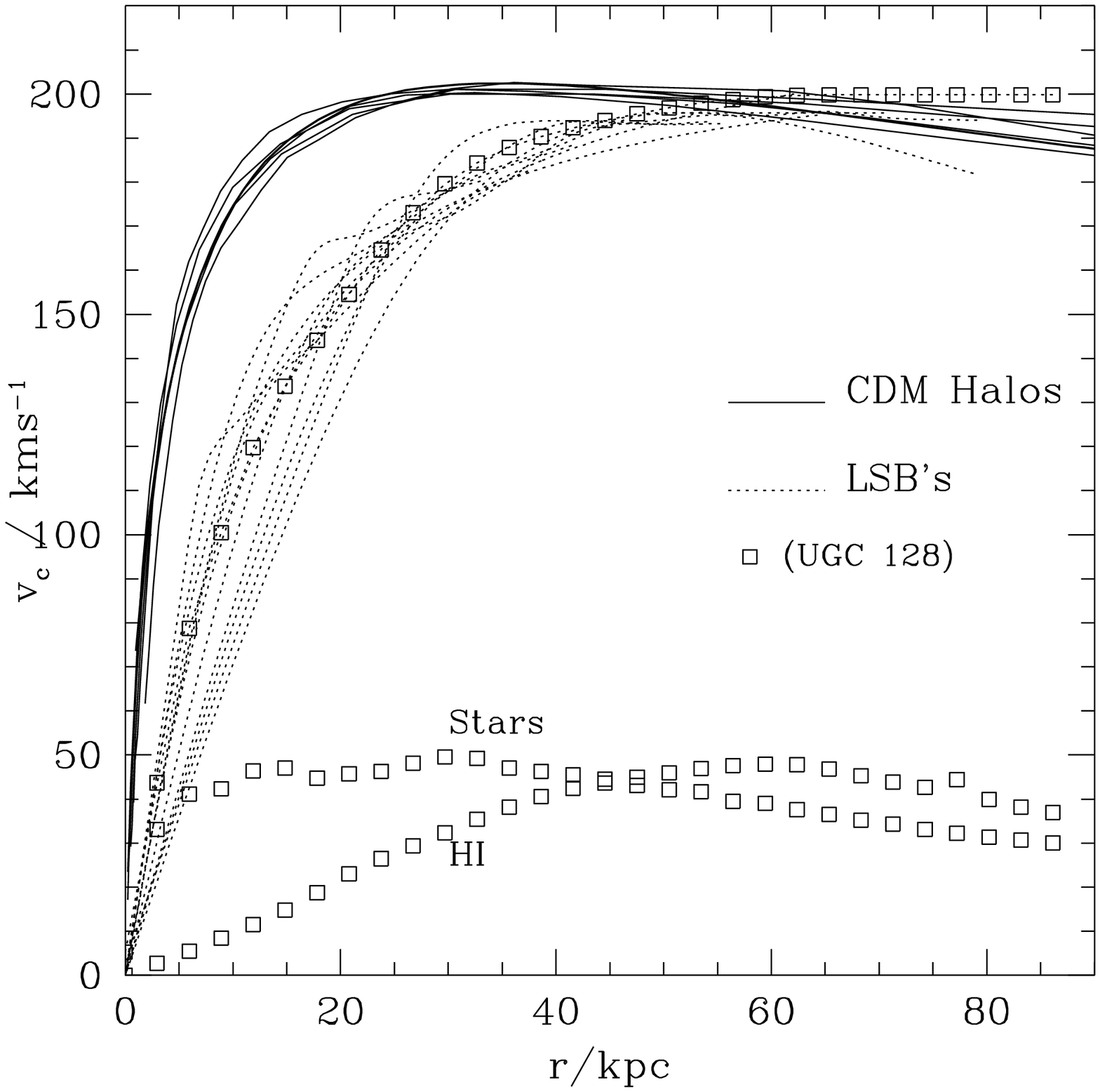}}

\noindent{\bf Figure 1.} \ {Rotation curves of high resolution CDM halos (solid curves)
compared with LSB rotation curve data (dotted curves). All of the data and model rotation
curves have been scaled to a fiducial peak velocity of $200\kms$. (Note that the simulation
halos and the data were chosen to have peak rotational velocities within 50\% of this value.)
The total rotational
velocity and the baryonic contribution from the stars and gas from a ``typical'' LSB
galaxy (UGC 128) are shown by the open squares.  The mass to baryon ratio for this galaxy
is nearly 20:1 and the rotation curve data probes a remarkable 25\% of the expected
virialised halo.

\

\

Figure 1 demonstrates that the standard CDM model fails to reproduce the rotation curves
of this sample of galaxies. Lowering the mass density will not solve this problem since
halos of a given mass would have larger concentrations since the mean density 
of the universe is higher during the collapse epoch.
Navarro \etal (1996) find that a flat Universe dominated by a lambda
term leads to halos with lower concentrations.
However, since the form of the density profile (and therefore the rotation curve) is
independent of cosmological parameters, we follow Kravtsov \etal (1998) and perform a 
model independent rotation curve shape comparison between simulations and data.

The shape of rotation curves of dark matter
dominated galaxies is universal across a wide range of galaxy luminosities (
Burkert 1995, Persic \& Salucci 1997, Kravtov \etal 1998).  Kravtsov \etal (1998)
fit a sample of dark matter dominated galaxies to the density profile
$$\rho(r)={{\rho_o}\over{(r/r_s)^\gamma [1+(r/r_s)^\alpha]^{(\beta-\gamma)/\alpha}}} \ , 
\ \ \ \ \ \ \ \ \ \ \ \ \ \ \ \ \ \ \ \ \ \ \ \ \ \ \ \ \ \ \ \ \ \ \ \ \ \ \ \ (1)$$ 
in order to obtain characteristic scale lengths $r_s$, and
circular velocities $v_s$ at $r_s$.  We reproduce their data for dwarf and LSB galaxies in
Figure 2.  Although the slope of the density profile in the outer regions is not
constrained by these data a good fit can be obtained using a constant density central
region i.e. $\alpha=2$,  $\beta=3$ and $\gamma=0$.
This is indicated by the solid curve that passes through the data points.

We find that the density profiles of our high resolution halos 
are well fit by a density model in which 
$\alpha=\gamma=1.5$ and $\beta=3$; {\it i.e.}
$$\rho(r) = {{\rho_o}\over{{[(r/r_s)^{1.5}(1+(r/r_s)^{1.5})]}}} \ , 
\ \ \ \ \ \ \ \ \ \ \ \ \ \ \ \ \ \ \ \ \ \ \ \ \ \ \ \ \ \ \ \ \ \ \ \ \ \ \ \ (2)$$ 
This profile is plotted in Figure 2, normalised to produce the same asymptotic flat slope
as the data points as well an additional curve stretched to provide a better fit to the
inner data.  Clearly, however we rescale this profile we cannot reproduce the slowly
rising rotation curve data.  The only caution that we can offer 
is that these data are all measured using HI. It would be very
difficult to measure these rotation curves using stellar spectroscopy, however, it is very
important to verify that the stars are tracing the same potential as the HI.

\centerline{\epsfysize=3.0truein \epsfbox{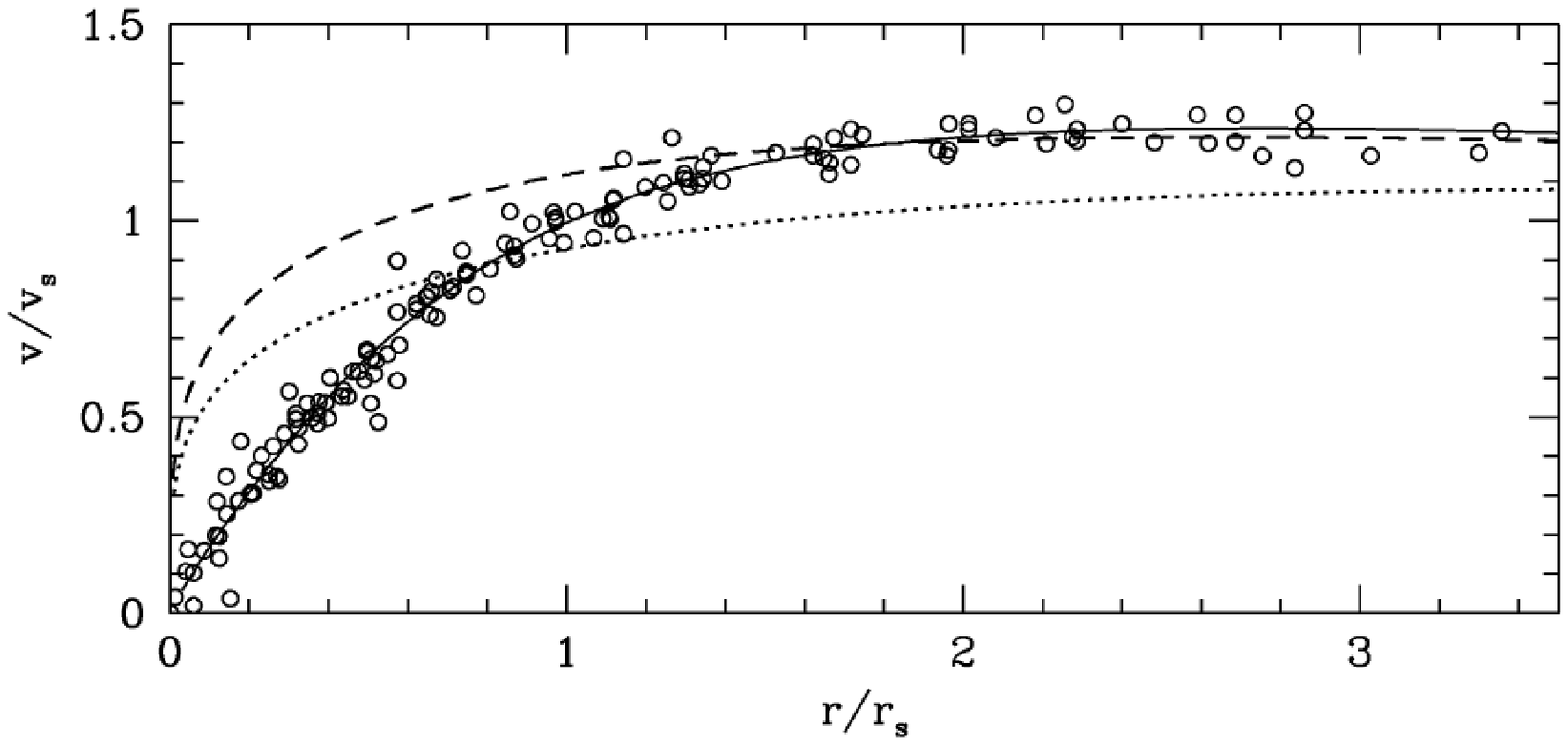}}

\noindent{\bf Figure 2.} \ The rotation curves of galaxies, from dwarf spirals
to giant LSB galaxies, scaled to fit the universal density profile of
equation 1. The solid curve shows the rotation curve that results from a
density profile with a constant density core, steepening to a slope of -3
at large radii.
The dashed and dotted curves show the rotation curves that result from
the density profiles of high resolution CDM halos normalised in two different ways.

\

\

Figure 3 shows the density profiles of our 2 highest resolution galaxy halos with circular
velocities $\sim 220 \kms$, compared with a CDM cluster that has a comparable resolution
(Moore \etal 1998).  
The density profiles of our halos can be fitted to within a few percent of the
data using equation 2, such that the central profile has a slope -1.5, curving to a slope
of -3 at the virial radius.  Here, $r_s=r_{vir}/c_m$, where for our fiducial galaxy halo
we find that a concentration $c_m=10$ fits at all resolved radii
({\it i.e.} beyond 2 softening lengths).  The cluster mass halo plotted in Figure 2 has a
virial mass $m_{vir}=4.3\times 10^{14}M_\odot$ within a virial radius $r_{vir}=1950$
kpc. The peak circular velocity of the cluster $v_{peak}=1100\kms$ which falls to
$v_{vir}=970\kms$ at the virial radius. Using equation 2 we find that a concentration
$c_m=4.0$ provides a good fit to the density profile.

The NFW density profile ($\alpha=\gamma=1$ and $\beta=3$) has a central slope of -1
curving more gently to the asymptotic slope of -3 at the virial radius. If we re-define
the concentration such that in equation 1, $r_s=r_{vir}/c_{nfw}$, we find that for our
fiducial galaxy halo a concentration $c_{nfw}=18$ fits our N-body data to within 20\%.
This value of the concentration is 50\% higher than the ``mean'' value predicted by the
procedure outline in Navarro \etal (1996).

The residuals from the data have a characteristic ``$S$'' shape; underestimating $\rho(r)$ at
the halo center and overestimating $\rho(r)$ in the mid range.  This value for the
concentration in the NFW profile gives a slope of -1.2 at 1 kpc, at which point the
profile in equation 1 has a slope of -1.6.  The ``average'' NFW fit to our fiducial halo
of this mass should
have a concentration $c=12$. In all cases we find that the best fit NFW profile requires a
concentration that is $50\%$ larger than the predicted value.  This may seem
unnecessarily pedantic, but determining the correct central density profile is essential
for comparing with observations that probe the inner few percent of the halo. This is not
just an important issue for a comparison with galactic rotation curves.  For example, the
gamma ray flux from neutralino-neutralino annihilation in the galactic center is an order
of magnitude higher when we use the profiles from our simulated halos 
rather than that predicted by NFW (Calcaneo-Roldan \& Moore 1999).

\centerline{\epsfysize=4.0truein \epsfbox{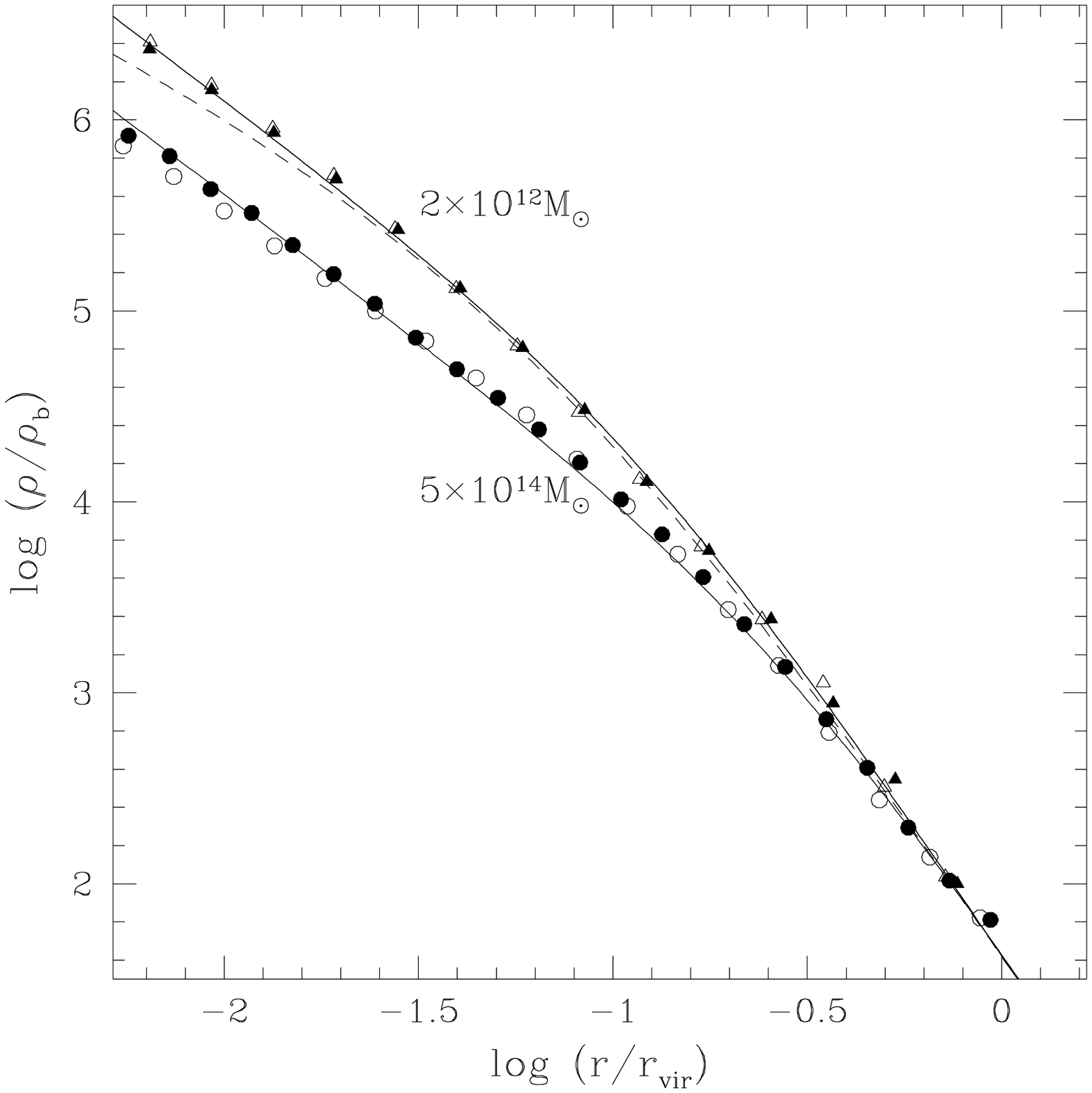}} 

\noindent{\bf Figure 3.} \ Density profiles of CDM halos on different mass scales.  Open
and closed triangles are two similar mass galactic halos whilst the filled circles show a
cluster mass halo. The open circles show the same cluster halo, but simulated with no
power below scales of 8 Mpc. The solid curves show fits to the data using equation 1 with
$c_m=10$ for the galaxies and $c_m=4$ for the cluster.  The dashed curve indicates an NFW 
profile fit to the galaxy data for points beyond 3\% of the virial radius, in this 
case $c_{nfw}=15$. 

\

\

\subsection{The effects of the merger history on halo density profiles}

Navarro \etal found that cluster mass halos have lower concentrations than
galaxies. i.e. at a fixed fraction of their virial radius, galaxies are denser than
clusters. We also find such a scaling. For comparison with the galactic halos, we plot one
of the cluster simulations from Moore \etal (1998) in Figure 3.  Navarro \etal suggest
that clusters are less concentrated than galaxies because they collapse at late times when
the universe is less dense.  In their scaling model, they take the redshift of collapse as
the time when half the mass of the final halo lies in progenitor clumps more massive than
a few percent of the final halo's mass.

The results of NFW have stimulated a great deal of theoretical work attempting
to explain both the shape of the dark matter density profiles and the scaling with
mass and cosmology. Modifications to the self similar collapse models 
({\it i.e.} Hoffman \& Shaham 1985) to include more realistic dynamics of
the growth process have been attempted 
({\it c.f.} Avila-Reese \etal (1998), 
Henricksen \& Widrow 1998, Lokas 1999, Kull 1999, 
Subramanian \& Ostriker 1999). 
Several authors argue that the central density profile is linked to the accretion and
merging history of dark matter substructure (Syer \& White 1998, Salvador-Sole \etal 1998,
Nusser \& Sheth 1999).  In order to test the importance of the merger history on halo
structure we perform the following experiment. We simulate the evolution of a CDM halo at
high resolution using the standard procedure outlined in Section 2. We then re-simulate
the same halo, but instead of using the full CDM power spectrum extrapolated to small
scales, we truncated the spectrum at a scale about half the size of the turnaround region
$\equiv 8$ Mpc.

Although the amplitude of fluctuations is similar on larger scales, the lack of small
scale power causes the halo to form via a single monolithic collapse rather than through
mergers and accretions of smaller halos.  This collapse is identical to that taking
place in a warm dark matter universe.  In this case the present day free-streaming
velocity is negligible and would not provide an additional phase space constraint on the
final core radius (Tremaine \& Gunn 1978). 

The distribution of mass at redshifts z=5 and z=0 are compared for these two simulations
in Figure 4 and the final density profiles are both plotted in Figure 3. The virial radii
of the two halos are the same, however, the lack of small scale power is clearly evident
in Figure 4.  Remarkably, the final density profiles are also very similar - the halo that
formed with no small scale power, has a slightly shallower central slope, -1.4 rather than
-1.5. The details of the merger history do not affect the final density profiles.

The concentration of a halo is most likely related to its collapse time, however the
definition of this epoch must be chosen with some care.  If we define the collapse time as
the epoch at which 90\% of the mass is in place, then both the halos simulated in Figure 4
form at a similar epoch. Defining the collapse time as the epoch when half of the final
mass is in smaller collapsed objects is not applicable in this case; 
at no time is the mass of the second simulation in objects much smaller 
than the final mass.

Complementary results were also found by Huss \etal (1999), who performed a series of
tests including isolated halo collapses, hierarchical collapses, varying amounts of
substructure and angular momentum. It appears that cuspy central density profiles are a
fundamental property of cold gravitational collapse. Neither the amount of small scale
power or the merger history can create significant changes in the halo density profiles.
In order to reproduce the observational data we must appeal to a new physical process that
leads to constant density cores. Such a mechanism is not inherent to our standard picture
of halo formation in cold or warm dark matter cosmologies.

Several authors have investigated the structure of galaxies that form from a
cold gravitational collapse of an isolated perturbation ({\it c.f.} van Albada
1982, Aguilar \& Merritt 1990).  These simulations typically produced good
agreement between the final stellar configuration and the de Vaucouleur
$r^{1/4}$ profiles. In 3 dimensions this profile has a significantly shallower
slope than $r^{-1}$, and may demonstrate the difference between cold
gravitational collapse using cosmological and isolated initial conditions.
However, it would be interesting to repeat these tests using the resolution that
can be achieved with current resources.

\

\centerline{\epsfysize=5.0truein \epsfbox{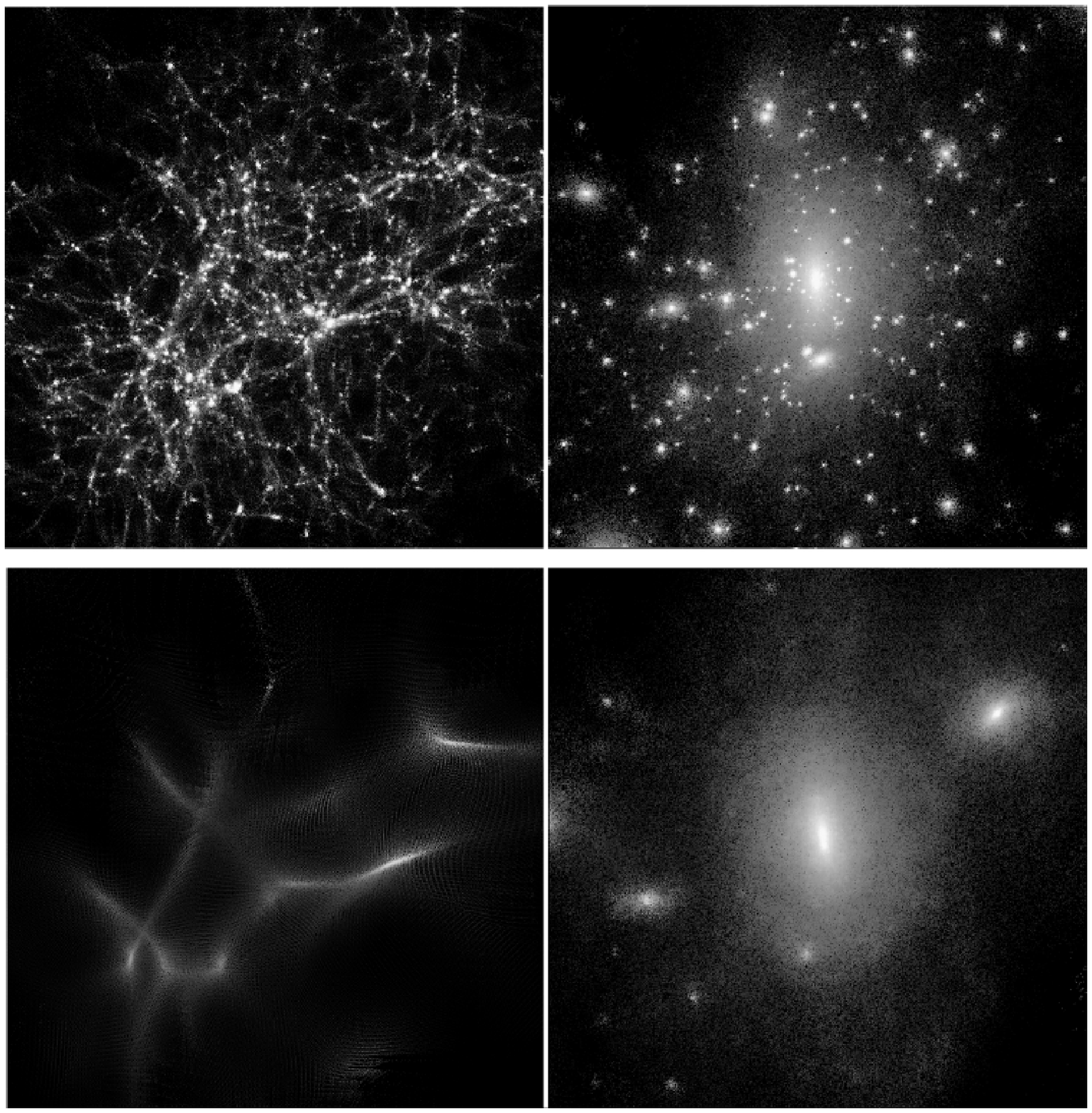}}

\noindent{\bf Figure 4.} \ The density field at z=5 (left panels) and z=0 (right
panels). The upper plots show the standard CDM halo simulation, the final density profile
is plotted as filled circles in Figure 2. This cluster is the same as analysed by Ghigna
\etal (1998).  The scale of the left hand panel is 20 Mpc across,
about twice the turnaround region, whereas the right hand panel is 8 Mpc and shows
the particles within twice the final virial radius.
The lower panels show the mass density field at the same epochs and on the same scales, 
but a cut-off has been placed in the power spectrum. 
The final density profiles of these halos are both plotted in Figure 3. 

\

\

\section{Summary and discussion}

We have performed a series of high resolution N-body simulations of cold dark matter halos
to compare with observations of galactic rotation curves.  Our resolution is sufficient to
resolve the expected rotation curves over the regions probed by the baryonic component. We
find a small scatter in the {\it central} 
density profiles which have asymptotic profiles with a
slope $\rho(r) \propto -1.5$.  Recent analysis of a CDM halo simulated with $10^7$ particles
and a resolution of 0.0005$r_{virial}$ demonstrates that this profile has converged 
(in preparation).

It is already established that CDM cannot provide the dark matter surrounding dwarf
galaxies since their rotation curves rise more slowly than earlier simulations suggested
(Moore 1994, Flores \& Primack 1994, Navarro \etal 1996). Recent proposals to alleviate
this problem have included explosive feedback (Navarro \etal 1997, Gelato \& Sommer-Larson
1998) or reducing the amount of CDM by adding a dark baryonic component (Burkert \& Silk
1997).

Low surface brightness galaxies provide the ideal systems for testing the structure of
dark matter halos. They span a range of masses and their luminous components are dominated
by dark matter over a large range of radii. The potentials are deep, therefore feedback,
however improbable, cannot be invoked as a means of forming central cores. In any case,
these galaxies have ``normal'' baryon fractions (at least as many baryons as HSB galaxies)
and they lie on the Tully-Fisher relationship.  Using a compilation of rotation curves
provided by Pickering \etal (1997) and de Blok \etal (1996) we find that CDM rotation
curves rise much too steeply to explain the observational data. Galaxies, such as UGC 128,
sample the halo mass distribution to a distance $\sim 25\%$ of the virial radius.  Their
rotation curves rise slowly within the inner 10\---20 kpc, and their rotation curves are
best fit with a large inner region that has close to a constant density.

These galaxies are not rare and they have properties that connect smoothly into samples of
higher surface brightness galaxies. Their surface brightness may result from larger than
average angular momentum in the dark matter component.  Since
angular momentum is uncorrelated with both environment and halo structure, we can
generalise our results to rule out the possibility that the dark matter in galactic halos
is a cold collisionless particle. Adjusting the cosmological parameters will not
change these conclusions since the shape of the density profile is independent of cosmology
and this shape is inconsistent with the observed HI rotation curves. However, these
results should be confirmed using stellar velocities of LSB disk stars.

CDM in its current incarnation, cannot provide the unknown mass that surrounds
galaxies. The dark matter must have an additional property that gives rise to the
core\--halo structure observed in LSB galaxies. We have shown that cutting the power
spectrum such that halos form from a single monolithic collapse would not change these
conclusions, therefore we cannot appeal to warm dark matter models.  A phenomenological
model that could reproduce the observations would be to invoke a dark matter particle with
a large cross-section for annihilation into ``hot dark matter'.  The mass annihilation
rate varies as $\rho^2 \sigma$, where $\sigma$ is the central velocity dispersion. The
annihilation by-products would stream from the halo centers leaving a core radius that
correlated directly with the circular velocity modulo a factor proportional to the
collapse time. This would also resolve the problem of the over-abundance of dark matter
substructure halos in CDM models since these would suffer early tidal disruption once
within the virial radius.

\acknowledgments

{\bf Acknowledgments} \ \ \ We would like to thank all of our colleagues for stimulating
discussions on dark matter halos.  We also thank Stacy McGaugh and Tim Pickering for
providing rotation curve data and comments and Andrey Kravtsov for providing additional
data for Figure 3.  The numerical simulations required many cpu hours, which were
primarily obtained as part of the Virgo consortium.  BM is a Royal Society Research
Fellow.

\baselineskip=8pt


\

\noindent{\bf References}





\pp Aguilar, L.A. \& Merritt, D. 1990, {\it Ap.J}., {\bf 354}, 33.

\pp Avila-Reese, V., Firmani, C. \& Hernandez, X. 1998, {\it Ap.J.}, {\bf 505}, 37.

\pp Brainerd, T.G., Goldberg, D.M. \& Villumsen, J.V. 1998, {\it Ap.J.}, {\bf 502}, 505.

\pp Burkert, A. 1995, {\it Ap.J.Lett.}, {\bf }, 447, L25

\pp Burkert, A. \& Silk, J. 1997, {\it Ap.J.Lett.}, {\bf 488}, L55.

\pp Calcaneo-Roldan, C \& Moore, B. 1999, in preparation.

\pp Carlberg, R. 1994, {\it Ap.J.}, {\bf 433}, 468.

\pp Cole, S. \& Lacey, C. 1996, {\it M.N.R.A.S.}, {\bf 281}, 716.



\pp Crone, M.M., Evrard, A.E. \& Richstone, D.O. 1994, {\it Ap.J.}, {\bf 434}, 402.

\pp Davis, M., Efstathiou, G., Frenk, C.S. \& White, S.D.M. 1985, 
{Ap.J.}, {\bf 292}, 371.

\pp de Blok, W.J.G., McGaugh, S.S. \& Van der Hulst, J.M. 
1996, {\it M.N.R.A.S.}, {\bf 283}, 18.

\pp de Blok, W.J.G. \& McGaugh, S.S. 1997, {\it M.N.R.A.S.}, {\bf 290}, 533.

\pp Dubinski, J. \& Carlberg, R. 1991, {\it Ap.J.}, {\bf 378}, 496.

\pp Eisenstein, D.J. \& Loeb, A. 1997, {\it Ap.J.}, {\bf 459}, 432.


\pp Evans, W.N. \& Collet, 1997, {\it Ap.J.Lett.}, in press.

\pp Flores, R.A. \& Primack, J.R. 1994, {\it Ap.J.Lett.}, {\bf 457}, L5.

\pp Frenk, C.S., White, S.D.M., Davis, M. \& Efstathiou, G. 1988, {\it Ap.J.}, 
{\bf 327}. 507.

\pp Fukushige, T \& Makino, J. 1997, {\it Ap.J.Lett.}, {\bf 477}, L9.

\pp Gelato, S., Sommer-Larson, J. 1999, {\it M.N.R.A.S.}, in press.

\pp Ghigna, S., Moore, B., Governato, F., Lake, G., Quinn, T. \& Stadel, J.
1998, {\it M.N.R.A.S.}, {\bf 300},  146.

\pp Henriksen, R.N \& Widrow, L.M. 1999, {\it M.N.R.A.S.}, {\bf 302}, 321.

\pp Hoffman, Y. \& Shaham, J. 1985, {\it Ap.J.}, {\bf 297}, 16.

\pp Huss, A., Jain, B. \& Steinmetz, M. 1999, {\it Ap.J.}, in press.

\pp Jing, Y.P. 1999, {\it Ap.J.}, submitted.

\pp Kull, A. 1999, {\it Ap.J.Lett.}, in press.

\pp Kravtsov, A.V., Klypin, A.A., Bullock, J.S. \& Primack, J.R. 1998, {\it Ap.J.},
{\bf 502}, 48.

\pp Lokas, E.L. 1999, {\it M.N.R.A.S.}, submitted.

\pp Moore, B., 1994, {\it Nature}, {\bf 370}, 620.

\pp Moore, B., Governato, F., Quinn, T., Stadel, J. \& Lake, G. 1998, 
{\it Ap.J.Lett.}, {\bf 499}, L5.


\pp Navarro, J.F., Frenk, C.S. \& White, S.D.M. 1996, {\it Ap.J.}, {\bf 462}, 563.

\pp Navarro, J.F., Eke, V.R. \& Frenk, C.S. 1996, {\it M.N.R.A.S.}, {\bf 283}, L72.

\pp Navarro, J.F. 1998, {\it Ap.J.}, submitted.

\pp Nusser, A. \& Sheth, R. 1998, {\it M.N.R.A.S.}, submitted.

\pp Peebles, P.J.E. 1984, {\it Ap.J.}, {\bf 277}, 470.

\pp Persic, M. \& Salucci P. 1997, {\it A.S.P.}, {\bf 117}, ed. Persic, M. and
Salucci P., 1.

\pp Pickering, T.E., Impey, C.D., Van Gorkom, J.H. \& Bothun, G.D. 1997,
{\it A.J.}, {\bf 114}, 1858.

\pp Quinn, P.J., Salmon, J.K. \& Zurek, W.H. 1986, {\it Nature}, {\bf 322}, 329.

\pp Quinn, T., Stadel, J. \& Lake, G. 1998, {\it Ap.J.}, submitted.

\pp Salvador-Sole, E., Solanes, J.M. \& Manrique, A. 1998, {\it Ap.J.}, {\bf 499}, 542. 

\pp Subramanian, K. \& Ostriker, J. 1999, in preparation.


\pp Syer, D. \& White, S.D.M. 1998, {\it M.N.R.A.S.}, {\bf 293}, 337.

\pp Tremaine, S. \& Gunn, J.E. 1978, {\it Phys.Rev.Lett.}, {\bf 42}, 407.

\pp Tormen, G., Bouchet, F.R. and White, S.D.M. 1996, {\it M.N.R.A.S.}, 
{\bf 286}, 865.

\pp Thomas, P. \etal 1998, {\it M.N.R.A.S.}, {\bf 296}, 1061.

\pp Van Albada, T.S. 1982, {\it M.N.R.A.S.}, {\bf 201}, 939.

\pp Warren, S.W., Quinn, P.J., Salmon, J.K. \& Zurek, H.W. 1992, {\it Ap.J.}, 
{\bf 399}, 405.

\pp Zwaan, M.A., van der Hulst, J.M., de Blok, W.J.G. \& McGaugh, S.S. 1995, 
{\it M.N.R.A.S.}, {\bf 273}, L35.

\baselineskip=14pt




\end{document}